\documentclass[aps,prd,preprint,onecolumn,amsmath,preprint,nofootinbib,nobibnotes,superscriptaddress]{revtex4}
\def\gsim{ \lower .75ex \hbox{$\sim$} \llap{\raise .27ex \hbox{$>$}} }
\def\lsim{ \lower .75ex \hbox{$\sim$} \llap{\raise .27ex \hbox{$<$}} }
\usepackage{graphicx,}

\begin{document}
\title{Dark Energy and Neutrino CPT Violation}
\author{Pei-Hong Gu}
\email{guph@mail.ihep.ac.cn} \affiliation{Theoretical Physics
Division, IHEP, Chinese Academy of Sciences, Beijing 100049, P. R.
China}
\author{Xiao-Jun Bi}
\affiliation{Key laboratory of particle astrophysics, IHEP,
Chinese Academy of Sciences, Beijing 100049, P. R. China}
\author{Xinmin Zhang}
\affiliation{Theoretical Physics Division, IHEP, Chinese Academy
of Sciences, Beijing 100049, P. R. China}

\begin{abstract}
In this paper we study the dynamical CPT violation in the neutrino
sector induced by the dark energy of the Universe. Specifically we
consider a dark energy model where the dark energy scalar
derivatively interacts with the right-handed neutrinos. This type
of derivative coupling leads to a cosmological CPT violation
during the evolution of the background field of the dark energy.
We calculate the induced CPT violation of left-handed neutrinos
and find the CPT violation produced in this way is consistent with
the present experimental limit and sensitive to the future
neutrino oscillation experiments, such as the neutrino factory.
\end{abstract}

\maketitle

\section{Introduction}
The recent observational data from type Ia
supernovae\cite{pelmutter}, cosmic microwave background (CMB)
radiation\cite{wmap} and large scale structure (LSS)\cite{sdss}
have provided strong evidences for a spatially flat and
accelerated expanding universe at the present time. In the context
of Friedmann-Robertson-Walker cosmology, this acceleration is
attributed to the domination of a component, dubbed dark
energy\cite{turner}. The simplest candidate for dark energy seems
to be a remnant small cosmological constant. However, many
physicists are attracted by the idea that dark energy is due to a
dynamical component, such as the quintessence
\cite{wetterich,ratra,frieman,zlatev}, the
K-essence\cite{armendariz,chiba}, the phantom\cite{phan} and the
quintom\cite{quintom,guo,xia,feng,li2}.

Being a dynamical component, the dark energy is expected to
interact with matters. Recently there have been a lot of interests
in the
literature\cite{li,gu,fardon,kaplan,bi,gu2,zhang,peccei2,bi2,horvat2,takahashi,fardon2,blennow,weiner,hli2,honda}
in studying the possible interactions between the neutrinos and
the dark energy. An interesting prediction of these models is that
the neutrino masses are not constant, but vary as a function of
time and space\cite{gu,fardon}. This prediction on the variation
of the neutrino masses can be verified in the present and future
experiments. For example, the neutrino mass evolution with time
can be tested in the measurement of the time delay of the
neutrinos emitted from the short gamma ray bursts\cite{hli}.
Another interesting possibility is to detect the neutrino mass
variation in space via the neutrino
oscillations\cite{fardon,kaplan,barger3,cirelli,barger4}. In this
paper we study the possibility of CPT violation in the neutrino
sector induced by the evolution of the dark energy scalar field.

In Ref.\cite{li} the authors have made a proposal for the
dynamical CPT violation by introducing a derivative coupling of
the dark energy scalar to the left-handed fermions of the standard
model\footnote{There are several other papers that deal with CPT
violation originating from spacetime-varying
scalars\cite{klp,blpr,kmst}}:
\begin{equation}
\mathcal{L}_{eff} \sim \partial_{\mu} \phi \bar{l}_{L} \gamma^\mu
l_{L},
\end{equation}
where $\phi$ is the dark energy scalar, for instance the
quintessence, $l_L$ by gauge invariance of the $SU(2)_L \times
U(1)_Y$ is the doublet of the left-handed lepton $l_L =(\nu_{L},
~e_{L})^{T}$. During the evolution of the universe the time
derivative of the scalar field does not vanish $\dot{\phi} \neq 0$
which gives rise to a CPT violation. This type of CPT violation as
shown in \cite{li} helps understand the matter anti-matter asymmetry
of the universe; however, since the laboratory experimental limit on
the CPT violation in electrons is so stringent that the induced CPT
violation in the neutrino sector will be much below the sensitivity
for the current and future experiments. Phenomenologically, a
comparably large neutrino CPT violation is helpful to explain the
possible CPT-violating neutrino oscillations. For example, the
neutrino CPT violation has been proposed
\cite{murayama,barenboim,barger,auerbach} to account for the LSND
anomaly\cite{lsnd}, which can be tested at the upcoming MiniBooNE
experiment\cite{miniboone}. It is noteworthy that whether the LSND
result is confirmed or not by the future experiments, the
possibility of CPT violation consistent with the other neutrino
oscillation experiments\cite{bahcall,conzalez-garcia,datta} is
intriguing and can be tested in the future neutrino factory
experiment\cite{barger2,datta}.

In this paper we consider specifically a model where the dark
energy scalar couples derivatively to the right-handed neutrino
and calculate the induced CPT violation in left-handed neutrinos
due to the mixing between the right-handed neutrinos and the
left-handed neutrinos. Our results show that with an appropriate
choice of the model parameter the CPT violation in the neutrino
sector is consistent with the present experimental data and can be
tested in the future neutrino factory experiment without
conflicting to the experimental limit on the electron CPT
violation.

\section{Model of neutrino CPT violation }

The model under investigation in this paper includes a scalar field
$\phi$ which drives the universe acceleration and derivatively
couples to the right-handed neutrinos $N_{i}=\nu_{Ri}+ \nu_{Ri}^{C}
(i=1, \cdot \cdot \cdot) $, $\sim \frac{f_{ij}}{\Lambda}
\bar{\nu}_{Ri}\partial _{\mu} \phi \gamma ^{\mu}\nu_{Rj}$ with $f$
the parameter which charactrizes the strength of this type of
interaction and is expected to be of order $\mathcal{O}(1)$, and
$\Lambda$ the energy scale charactrizing the dynamics which
generates this effective interaction. In our model, the right-handed
neutrinos, being a part of the dark energy sector, are taken at the
$\mathcal{O}( \rm eV )$ scale and $\Lambda \ll M_{W}$ to give
observable CPT violating effects. Similar $\mathcal{O}( \rm eV )$
scale right-handed neutrino models are studied in
\cite{fardon,kaplan}. The derivative coupling will lead to a
cosmological CPT violation in the right-handed neutrino sector
during the evolution of the background dark energy field. The mixing
between the left-handed neutrino and the right-handed neutrino is
induced through a gauge invariant and renormalizable Yukawa
interaction, $y_{\alpha j}\bar{l}_{L \alpha} \tilde{H} \nu_{Rj}
(\alpha =e,\mu,\tau \textrm{ denotes the flavor indices})$, where
$y$ are the Yukawa couplings, $l$ and $ \tilde{H}$ are the SM lepton
doublet and Higgs doublet, respectively.

The relevant Lagrangian can be written as
\begin{displaymath}
-\mathcal{L}=y_{\alpha j}\bar{l}_{L \alpha} \tilde{H}
\nu_{Rj}+\frac{1}{2}M_{i}\bar{\nu}_{Ri}^{C}
\nu_{Ri}+h.c.+\frac{f_{ij}}{\Lambda} \bar{\nu}_{Ri}\partial _{\mu}
\phi \gamma ^{\mu}\nu_{Rj}-\frac{1}{2} \partial _{\mu}\phi
\partial ^{\mu}\phi +V(\phi)
\end{displaymath}
\begin{equation}
\label{lag} =y_{\alpha j}\bar{l}_{L \alpha} \tilde{H}
N_{j}+h.c.+\frac{1}{2}M_{i}\bar{N}_{i}
N_{i}+\frac{1}{2}\bar{N}_{i}A_{\mu ij} \gamma ^{\mu}
N_{j}+\frac{1}{2}\bar{N}_{i}S_{\mu ij} \gamma ^{\mu}\gamma _{5}
N_{j}-\frac{1}{2} \partial _{\mu}\phi \partial ^{\mu}\phi +V(\phi)
\end{equation}
with $V(\phi)$ the potential for $ \phi $. We have adopted the
following definition in Eq.(\ref{lag}),
\begin{equation}
\label{matrix} A_{\mu}=\frac{1}{2}(f-f^{T}) \frac{1}{\Lambda}
\partial _{\mu} \phi\ ,
\end{equation}
\begin{equation}
S_{\mu}=\frac{1}{2}(f+f^{T}) \frac{1}{\Lambda} \partial _{\mu}
\phi\ ,
\end{equation}
so the matrix $A_{\mu}$ is antisymmetric while $S_{\mu}$ is
symmetric.

\begin{figure}
\includegraphics[scale=1.0]{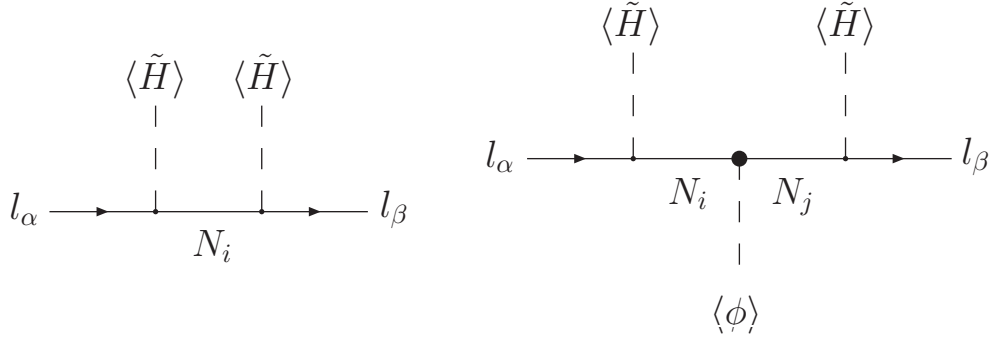}
\caption{Seesaw mechanism (the left diagram) and CPT violation in
neutrinos (the right diagram) at tree level.}
\end{figure}

Integrating out the right-handed Majorana neutrinos, we obtain
\begin{equation}
\label{cptv}
-\mathcal{L}_{\nu}=\frac{1}{2}\bar{\nu}_{\alpha}m_{\nu}^{\alpha
\beta} \nu_{\beta}+ \frac{1}{2}\bar{\nu}_{\alpha}a_{\mu}^{\alpha
\beta} \gamma ^{\mu} \nu_{\beta}+ \frac{1}{2}
\bar{\nu}_{\alpha}s_{\mu}^{\alpha \beta} \gamma ^{\mu} \gamma _{5}
\nu _{\beta}\ .
\end{equation}
with
\begin{equation}
\label{seesaw} m_{\nu}=-m_{D}\frac{1}{M}m_{D}^{T}\ ,
\end{equation}
\begin{equation}
a_{\mu}=\frac{1}{2}(b_{\mu}-b_{\mu}^{T})
\end{equation}
and
\begin{equation}
s_{\mu}=-\frac{1}{2}(b_{\mu}+b_{\mu}^{T})\ ,
\end{equation}
where $\nu =\nu_{L}+ \nu_{L}^{C}$ is the left-handed Majorana
neutrino. Here $M=\textrm{diag}(M_{1},M_{2},...)$, $m_{D}=yv$ with
$v \simeq 174 \textrm{GeV}$ and  $b_{\mu}$ is defined as
\begin{equation}
\label{bmu} b_{\mu}=m_{D} \frac{1}{M}(A_{\mu}-S_{\mu})
\frac{1}{M}m_{D}^{\dagger}\ .
\end{equation}
We can see that the last two terms in Eq.(\ref{cptv}) will produce
the CPT violation in neutrinos for a non-vanishing $\dot\phi$. It is easy
to show that the CPT
violating term above can be given in a simple
form\cite{barger2,kostelecky,colladay,coleman}
\begin{equation}
\label{form} \mathcal{L}_{CPTV}=-\bar{\nu}_{L
\alpha}b_{\mu}^{\alpha \beta} \gamma ^{\mu} \nu_{L \beta}\ .
\end{equation}

For the electron the CPT violation is induced by the W- and
neutrino-loop\footnote{The $\tilde{H}-N$ exchange also contributes
to the CPT violation in electrons and neutrinos\cite{mocioiu}, which is,
however, strongly suppressed in our model since the masses of the
right-handed Majorana neutrinos and also the Yukawa couplings are
very small.}. The coupling between the axial vector background and
the electron current, $c_{\mu} \bar{e} \gamma ^{\mu} \gamma
_{5}e$, is strongly constrained by the present
experiments\cite{adelberger}. For $ \Lambda \ll M_W$, we obtain
\begin{equation}
\label{cmu} c_{\mu} \simeq \frac{1}{8 \pi} \frac{\alpha}{ \sin
^{2}\theta_{W}} \frac{\Lambda
^{2}}{M_{W}^{2}}s_{\mu}\ ,
\end{equation}
where the loop integral was cutoff at $\Lambda$ as the effective theory
is only valid below $\Lambda$ \cite{mocioiu},
$\alpha \simeq \frac{1}{137}$ is the fine structure constant,
$ \sin ^{2} \theta_{W} \simeq 0.23$ is the Weinberg angle, and
$M_{W} \simeq 80 \textrm{GeV}$ is the mass of $W$ gauge boson.

\begin{figure}
\includegraphics[scale=1.0]{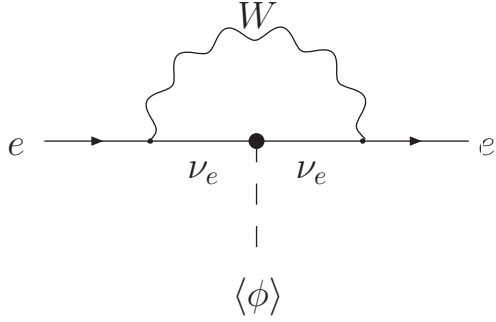}
\caption{CPT violation in electrons at one loop level.}
\end{figure}

Taking the CPT violating parameters $a_{\mu}$ and $s_{\mu}$ to be
much smaller than the masses of the left-handed Majorana
neutrinos, which will be shown latter to be the actual case, we
can simplify Eq.(\ref{bmu}) as
\begin{equation}
\label{bmu2} b_{\mu} \sim \tilde{b} \frac{m}{M}\frac{1}{\Lambda}
\partial _{\mu}\phi
\end{equation}
and similarly for Eq.(\ref{cmu}) we have
\begin{displaymath}
c_{\mu} \sim \tilde{c} \frac{1}{8 \pi} \frac{\alpha}{ \sin
^{2}\theta_{W}} \frac{\Lambda^{2}}{M_{W}^{2}} \frac{1}{\Lambda}
\partial _{\mu}\phi
\end{displaymath}
\begin{equation}
\label{cmu2} \sim 1.3 \times 10^{-3} \tilde{c} \frac{\Lambda
}{M_{W}^{2}} \partial _{\mu}\phi\,,
\end{equation}
where $m$, $M$ denote the mass scale of the left- and right-handed
Majorana neutrinos, $\tilde{b}$ and $\tilde{c}$ are two
dimensionless parameters. It should be noted that the magnitude of
$\tilde{b}$ and $\tilde{c}$ are at the same order of $f \sim
\mathcal{O}(1)$ in Eq.(\ref{lag}).

For the homogeneous scalar field, we can express the time
component of
$\partial_{\mu}\phi$ in the following way,
\begin{equation}
\label{dotphi} \dot{\phi}=[(1+w_{\phi})\rho_{\phi}]^{1/2}
\end{equation}
with the equation of state defined by
\begin{equation}
\label{w} w_{\phi} =\frac{\frac{1}{2}
\dot{\phi}^{2}-V(\phi)}{\frac{1}{2} \dot{\phi}^{2}+V(\phi)}
\end{equation}
and the energy density
\begin{equation}\label{rhode}
\rho_{\phi} =\frac{1}{2} \dot{\phi}^{2}+V(\phi)\ .
\end{equation}
The values of $w_{\phi}$ and $\rho_{\phi}$ are constrained by the
cosmological observations\cite{wmap}, $w_{\phi}<-0.78$ and
$\rho_{\phi}\simeq 73\%\rho_{c}$ with $\rho_{c} \simeq 4.2 \times
10^{-47} \textrm{GeV}^{4}$ the critical energy density of the
Universe at present. Accordingly we have for $\dot{\phi}$
\begin{equation}
\label{dot2} \dot{\phi}\lesssim 2.6\times
10^{-24}\textrm{GeV}^{2}\ .
\end{equation}
The CPT violating parameters of the neutrinos and the electrons
are now given by
\begin{equation}
\label{b} b_{0} \sim
\tilde{b}\frac{m}{M}\frac{1}{\Lambda}\dot{\phi}
\end{equation}
and
\begin{equation}
\label{c} c_{0} \sim 1.3 \times
10^{-3}\tilde{c}\frac{\Lambda}{M_{W}^{2}}\dot{\phi}\
.
\end{equation}

\section{Experimental test for the CPT violation}
The neutrino CPT violation can be tested in the neutrino
oscillation experiments. So we begin our discussions with the
effective Hamiltonian that governs the propagation of neutrinos.
The evolution of neutrinos is determined by the Schr$
\ddot{o}$dinger equation
\begin{equation}
\label{sch} i \frac{d}{dt} \left(\begin{array}{cc}
\nu_{e} \\
\nu_{\mu} \\
\nu_{\tau}
\end{array}\right) \simeq H_{eff}(x) \left(\begin{array}{cc}
\nu_{e} \\
\nu_{\mu} \\
\nu_{\tau}
\end{array}\right),
\end{equation}
where the Hamiltonian can be written as\cite{kostelecky}
\begin{displaymath}
H_{eff}(x)=\frac{1}{2|\vec{p}|}(m_{\nu}m_{\nu}^{\dagger})+\frac{1}{|\vec{p}|}(b_{\mu} p^{\mu})+\sqrt{2}G_{F}\textrm{diag}(N_{e}(x),0,0)
\end{displaymath}
\begin{equation}\label{ham}
\simeq
\frac{1}{2p^{0}}(m_{\nu}m_{\nu}^{\dagger})+b_{0}-|\vec{b}|\cos
\theta + \sqrt{2}G_{F}\textrm{diag}(N_{e}(x),0,0)
\end{equation}
with $m_{\nu}$ and $b_{\mu}=(b_{0},-\vec{b})$ defined in
Eq.(\ref{seesaw}) and Eq.(\ref{bmu}), respectively.
$p^{\mu}=(p^{0},\vec{p})$ is the four momentum of the neutrinos,
$\theta$ is the angle between $\vec{p}$ and $\vec{b}$, and
$\sqrt{2}G_{F}N_{e}(x)$ is the usual MSW term\cite{msw}.

We define $b_{i}$ to be the eigenvalues of the matrix $b_{0}$ and
$\delta b_{ij}=b_{i}-b_{j}$. Then $\delta b_{21}$ is constrained
by fitting the data from the solar neutrino experiments and the
KamLAND reactor neutrino experiment\cite{bahcall}, which gives
\begin{equation}
\label{b21} \delta b_{21} < 3.1 \times 10^{-20} \textrm{GeV} \ .
\end{equation}
In addition, the bound of $\delta b_{32}$ is also obtained from
the Super-K and K2K data\cite{conzalez-garcia,datta}
\begin{equation}
\label{b32} \delta b_{32} < 5 \times 10^{-23} \textrm{GeV} \ .
\end{equation}
And it has been shown the
future neutrino factory experiment\cite{barger2,datta} will be sensitive
to $\delta b_{32}$ as small as
\begin{equation}
\label{deltab32} \delta b_{32} \sim 3 \times 10^{-23} \textrm{GeV}
\ .
\end{equation}

>From the cosmological observations\cite{wmap} and the neutrino
oscillation experimental results\cite{sk,sno,k2k,kamland}, we take
the parameters of our model as $\rho_{\phi} \sim 3\times 10^{-47}
\textrm{GeV}^{4}$ and $w_{\phi}\sim -0.9 $ and the mass scale of
the neutrinos as $m\sim 10^{-2} \textrm{eV}$. For the mass of
right-handed neutrino (dark fermion) we take for example
$M=1\textrm{eV}$\cite{fardon}. One can see the values of the CPT
violation parameters $\delta b_{21}$ and $\delta b_{32}$ are
predicted to be
\begin{equation}
\label{db21} \delta b_{21} \sim \delta {\tilde b_{21}}
(\frac{m}{M})\frac{1}{\Lambda}[(1+w_{\phi})\rho_{\phi}]^{1/2} \sim
\delta {\tilde b_{21}} (\frac{\rm MeV}{\Lambda}) 1.7\times
10^{-23} \textrm{GeV} \ ,
\end{equation}
and
\begin{equation}
\label{db32} \delta b_{32} \sim \delta {\tilde b_{32}}
(\frac{m}{M})\frac{1}{\Lambda}[(1+w_{\phi})\rho_{\phi}]^{1/2} \sim
\delta {\tilde b_{32}} (\frac{\rm MeV}{\Lambda}) 1.7\times
10^{-23} \textrm{GeV}\ .
\end{equation}
As shown in Eq.(\ref{bmu2}) the parameters $\delta {\tilde
b_{32}}$ and $\delta {\tilde b_{21}}$ are $O(1)$. The equations
above indicate that it is quite possible to have CPT violating
effects be consistent with the present experimental data and be
detectable in the future neutrino factory experiment with
$\Lambda$ no larger than MeV. It should be noted that if the value
of the quintessence field $\phi$ is at the order of $M_{Pl}$ the
mass of the right-handed neutrinos should be shifted greatly by
Eq. (\ref{lag}). However, in the dark energy
models\cite{fardon,khoury} where the value of $\phi$ is much lower
than $M_{Pl}$ even lower than $\Lambda$, our model is consistent.

We now estimate the corresponding CPT violation in the electron
sector. For the parameters taken above, we have $c^{0} \sim
\tilde{c}\ 3.5 \times 10^{-34} \textrm{GeV}$. The present
experimental limit is $c^{0} < 5 \times 10^{-25} \textrm{GeV}$
\cite{barger2,mocioiu}. We see the electron CPT violation in our
model is well within the present limit.

\section{Conclusion}
In summary, we have studied the neutrino CPT violation induced
from the dark energy sector. Specifically we consider a dark
energy model including a scalar boson and right-handed neutrinos,
and introduce a derivative coupling between the boson and the
neutrinos. This derivative coupling leads to a cosmological CPT
violation during the evolution of the dark energy field. We
calculate the induced CPT violation in the left-handed neutrinos
due to their mixing with the right-handed neutrinos, and also the
loop contribution to the electron CPT violation.
 Our calculations show that the CPT violation in both
neutrinos and electrons is well within the present experimental
limits. Furthermore, the neutrino CPT violation can be tested in
the future neutrino oscillation experiments, such as the neutrino
factory. Our study in this paper shows the neutrino oscillations
may provide us a non gravitational way to probe the dark energy
properties.

\begin{acknowledgments}
We thank Zhi-Hai Lin for helpful discussions. This work is
supported in part by the NSF of China under the grant No.
1057511, 10120130794, 10105004, 19925523, 90303004 and also by the
Ministry of Science and Technology of China under grant No. NKBRSF
G19990754.
\end{acknowledgments}

\end{document}